\documentclass[runningheads]{llncs}
\pdfoutput=1
\usepackage{graphicx}
\usepackage{subcaption}
\usepackage{amsfonts}
\usepackage[T1]{fontenc}

\begin{document}

\title{Quantifying Synchronization in a Biologically Inspired Neural Network}
\titlerunning{Synchronisation in a Bio-inspired Neural Network}

\author{Pranav Mahajan\inst{1} \and
Advait Rane\inst{2} \and
Swapna Sasi\inst{2} \and
Basabdatta Sen Bhattacharya\inst{2}}
\authorrunning{Mahajan et al.}

\institute{Department of Electrical and Electronics Engineering, BITS Pilani University, K. K. Birla Goa Campus, India \and
Department of Computer Science and Information Systems, BITS Pilani University, K. K. Birla Goa Campus, India\\
\email{basabdattab@goa.bits-pilani.ac.in}\\
\url{https://www.binnlabs-goa.in/}
}
\maketitle              
\begin{abstract}
We present a collated set of algorithms to obtain objective measures of synchronisation in brain time-series data. The algorithms are implemented in MATLAB; we refer to our collated set of ‘tools’ as SyncBox. Our motivation for SyncBox is to understand the underlying dynamics in an existing population neural network, commonly referred to as neural mass models, that mimic Local Field Potentials of the visual thalamic tissue. Specifically, we aim to measure the phase synchronisation objectively in the model response to periodic stimuli; this is to mimic the condition of Steady-state-visually-evoked-potentials (SSVEP), which are scalp Electroencephalograph (EEG) corresponding to periodic stimuli. We showcase the use of SyncBox on our existing neural mass model of the visual thalamus. Following our successful testing of SyncBox, it is currently being used for further research on understanding the underlying dynamics in enhanced neural networks of the visual pathway.


\end{abstract} 
\section{Introduction}
\label{sec:1}
Phase-locking behaviour in neural populations is hypothesised as the basis of efficient information processing in the visual cortex \cite{martin2016phase}. Phase synchronisation is defined as ``the synchronisation between chaotic systems taking only phase-locking into account, with no restriction on amplitudes'' ~\cite{Rosenblum2001}. Periodicity and synchrony are ubiquitous in the nervous system both in health (eg.\ sleep~\cite{Krueger2013}) and disease (eg.\ Parkinsonian tremor~\cite{tassetal1998physrevlett}). Thus, neuronal dynamics underlying rhythmic periodicity has been the subject of several research ~\cite{glass2001synchronization,longtin1994bistability,farokhniaee2017mode,tassetal1998physrevlett}. Of particular interest is the neuronal dynamics corresponding to Steady-state-visually-evoked-potentials (SSVEP), that refer to Electroencephalogram (EEG) recorded from humans and animals when subjected to rhythmic visual stimuli. For a review on the usefulness of SSVEPs in understanding visual information processing and transmission, as well as its application to Brain-Computer Interface, readers may refer to~\cite{norcia2015steady,Vialatte2010,Ku2013}. A proof-of-concept study have validated biologically inspired neural population networks, commonly referred to as neural mass models, to underpin causality in the visual brain circuit corresponding to SSVEP~\cite{maciej2016}. However, these studies rely on a qualitative evaluation of model response. Indeed, what is desirable in these studies is a set of robust methods that provide an objective measure of phase synchronisation between neural populations in response to periodic input. In this work, we have collated a set of tools implemented on MATLAB to serve this purpose. We refer to our set of tools as SyncBox.

SyncBox implements the following synchronisation measures: Phase Locking Value (PLV), Normalized Shannon Entropy (NSE), Mutual Information, Lambda Synchronization Index (LSI), Spectral Coherence and Phase slip. We have also added Phase histogram plots wrapped on the cyclic degree scale for easy visualisation of synchronisation in noisy time-series data. PLVs are used to explore under what conditions synchronisation take place in a system and is often used to inspect an Arnold Tongue formation of a system response ~\cite{Lachaux1999}. 

Spectral coherence has been widely used for measuring synchronisation between two signals and gives a fair idea of relevant coherent frequencies in data, ~\cite{Lowet2016}, however, PLV is thought to be a more true indicator of synchrony~\cite{kreuz2013synchronization}  \cite{Lowet2016}. Normalised Shannon Entropy and the Lambda Measure have been identified as useful n:m synchronisation indices for the statistical analysis of the phase difference between interacting oscillators~\cite{Rosenblum2001}. Phase slips and polar histogram plots provide a clear way to visualise the phase difference behaviour and put the above measures into perspective~\cite{Notbohm2016}.

We test our toolbox on an existing neural mass model of the visual thalamus Lateral Geniculate Nucleus (LGN)~\cite{bsb_springerbookchap_2016}. We present the model with periodic input at varied frequency and at varying input strength. Applying the various tools in SyncBox allows an objective understanding of the phase synchrony and phase locking in connected neuronal populations, as well as the causality with varying connectivity strengths. Upon successful validation, we are currently implementing this toolbox for further research.

The layout of our paper is thus: In Section~\ref{sec:2}, we provide a brief overview of the various measures implemented in SyncBox as well as of the LGN network. The simulation methods used in this work are also presented. In Section~\ref{sec:3}, we present and discuss the results of testing SyncBox on the LGN network response. We conclude the paper in Section~\ref{sec:4}.
 
\section{Materials and Methods}
\label{sec:2}

\subsection{Phase Locking Value}
\label{sec:21}

The phase relation (i.e. the generalized phase difference) between two coupled system is defined as follows,
\begin{equation}
\phi_{n,m} (t) = \mathopen|n \phi_1 (t) - m \phi_2 (t) \mathclose| 
\label{plv-eq1}
\end{equation}
And the condition for phase locking is as follows,
\begin{equation}
\mathopen|n \phi_1 (t) - m \phi_2 (t) - \delta \mathclose| < \textrm{constant} \; \forall t  \label{plv-eq}
\label{plv-eq2}
\end{equation}
 where, $t$ represents time and $\delta \in \mathbb{R}$ is some average phase shift (a constant value) and $\phi_{n,m} (t)$ is the instantaneous phase relation between the coupled systems \cite{Rosenblum2001}. In cases of perfect phase locking, the phase relation $\phi_{n,m} (t) = \delta$, but in realistic conditions of phase synchronization, $\phi_{n,m} (t)$ is nearly constant $\forall t$ and observes a dirac-delta like distribution or a "peaky" distribution, where $\phi_{n,m} (t)$ might oscillate $\delta$ (bounded by some constant value as shown in \ref{plv-eq2})\cite{Rosenblum2001}. $n,m \in \mathbb{N}$ represents the mode of phase locking $n:m$. They also represent frequency ratios $m:n$ in which coupled systems can satisfy the condition of phase locking despite different frequencies $\omega_1$ and $\omega_2$ as differentiating equation \ref{plv-eq1}, gives us $n\omega_1 \approx m\omega_2$ which is the frequency entrainment condition. For example, the modes of phase locking $n:m$ could be 1:1, 1:2, 2:1, 2:3 etc. PLV takes a value from 0 to 1; zero signifying no phase synchrony and one signifying complete phase synchrony.

\begin{figure}[h!]
\centering
\includegraphics[width=0.75\linewidth]{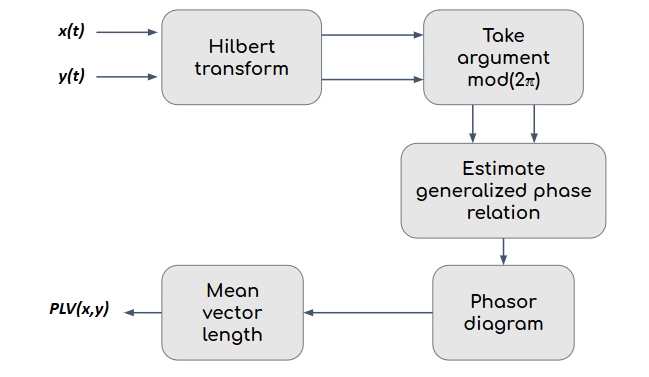}
\caption{Algorithm for calculating Phase Locking Value (PLV) presented as a signal-flow diagram.}
\label{fig:illustration2}
\end{figure}

To compute the PLV between two-time series  $x(t)$ and $y(t)$, the instantaneous phases of the signals at each time step are extracted using Hilbert transform. Instantaneous linear (unwrapped) phases are converted to wrapped phases which take values from $-\pi$ to $+\pi$. Then the phase relation at each time step is represented as a Phasor diagram. Mean vector length of all the phase relations at different time steps is computed as the PLV \cite{Rosenblum2001} (See Fig. \ref{fig:illustration2} for a schematic representation).

It is useful to obtain a global map of synchronization regions to analyze the synchronization in population models in response to periodic external stimuli. Synchronization between spiking time-series output of different neuronal populations or between the spike trains and an external input depends on both amplitude and frequency of the input. Thus, one can obtain regions of phase-locking on an amplitude-frequency plane, and these are commonly referred to as Arnold Tongue, due to the tongue like shape chalked out by the high synchrony regions \cite{farokhniaee2017mode}.

\subsection{Normalised Shannon Entropy}
\label{sec:23}
The Shannon Entropy (SE) of a distribution indicates the amount of uncertainty in the value of the variable \cite{shannon1948mathematical}. For the distribution given by the difference in the phase angles between two signals, SE will indicate the number of different values the phase difference takes. This value can be normalised as described below to obtain the Normalised Shannon Entropy (NSE), $\rho_{norm} \in [0, 1]$, as a measure of the synchronization between the two signals \cite{Rosenblum2001}. 

To give a qualitative description, the phase difference for perfectly synchronised signals, as we know, gives a Dirac-like distribution which has a low Shannon entropy. For independent signals with no synchronisation, the phase difference is free to take any value between $-\pi$ to $\pi$, giving a uniform distribution with high Shannon entropy \cite{Rosenblum2001}. The Shannon entropy is normalised to reflect synchronization.

We have used the implementation given by Notbohm et. al. \cite{Notbohm2016}. We take the Hilbert transform of both the signals and calculate the unwrapped phase difference between the angles of the transform. The phase difference is then wrapped between $-\pi$ and $\pi$, and binned as a histogram with $N=80$ bins, where $p_k$ gives the probability of the $k$th bin. The Shannon entropy, $S$ of this distribution is calculated and normalised to $\rho_{norm}$ using the following equations,

\begin{equation}
    S = -\sum_{k=1}^{N} p_k \ln p_k
\end{equation}

\begin{equation}
    \rho_{norm} = \frac{S_{max} - S}{S_{max}} 
\end{equation}

where $S_{max}$ is the maximum possible Shannon entropy given by $\ln N$. $\rho_{norm}$ is closer to one for synchronized signals and closer to zero for independent unsynchronized signals.

\subsection{Lambda Synchronisation Index}
\label{sec:24}
The Lambda Synchronisation Index (LSI) is an implementation of the synchronisation measure described in Rosenblum et al~\cite{Rosenblum2001}. The phase relation between two oscillating signals is quantified thus: the phase of the second oscillating signal, $\phi_2$, is observed at instants of time when the phase of the first oscillating signal, $\phi_1$, attains a certain fixed value $\theta$. $n:m$ phase locking is accounted for by wrapping the phases in intervals of $[0, 2\pi n]$ and $[0, 2\pi m]$ respectively. The observed phase of the second signal is represented by $\eta$.
\begin{equation}
    \eta = \phi_2 mod2\pi n|_{\phi_1mod2\pi m = \theta}
\end{equation}
The magnitude of the sum of the unit complex vectors given by the observed phase angles of the second signal is used as a measure of synchronisation. For perfectly synchronised signals where the phase difference is constant, $\eta$ would take up the same value for all instances. The sum of the unit complex vectors would be concentrated in one direction. On the other hand, for independent signals, $\eta$ would be free to take up all values between $-\pi$ and $\pi$. In this case, the unit complex vectors would cancel each other out. Let the phase of the first signal have $N$ possible values. For the $l$th value of the first phase, let the phase of the second signal have $M_l$ values. The averaged sum for each $l \in [1, N]$ is given by, 
\begin{equation}
    \Lambda_l = M_l^{-1}\sum_{i=1}^{M_l}\exp(i(\frac{\eta_{i,l}}{n}))
\end{equation}
Finally, we take the addition of the magnitude of all $N$ values of $\Lambda_l$. Thus, the LSI is given by,
\begin{equation}
    \lambda_{n,m} = N^{-1}\sum_{l=1}^N |\Lambda_l|
\end{equation}
Since $\phi_1$ can take infinitely many values, for the purpose of calculation the values of $\phi_1$ are binned into $N$ bins. For our implementation, we have calculated the optimal number of bins using the following formula as described in Rosenblum et al~\cite{Rosenblum2001},
\begin{equation}
    N = \exp[0.626 + 0.4ln(M - 1)]
\end{equation}
where M is the total number of samples to be binned.

\subsection{Spectral Coherence}
\label{sec:25}
Spectral coherence $C_{xx}$ measures the normalized correlation between two power spectra. 
\begin{equation}
C_{xy}(f) = \frac{|G_{xy}(f)|^2} {G_{xx}(f) G_{yy}(f)}
\label{coh-eqn}
\end{equation}

where $G_{xx}$ is the power spectral density (PSD) estimate of signal $x(t)$, $G_{yy}$ is the PSD of signal $y(t)$ and $G_{xy}$ is the cross-spectral density (CSD) estimate of signals $x(t)$ and $y(t)$ \cite{kreuz2013synchronization}\cite{Lowet2016}. CSD is calculated as the Fourier transform of cross-correlation of the signals $x(t)$ and $y(t)$. One may think of the CSD as measuring hidden periodicities in the cross-correlation, just like the PSD measures periodicities in the autocorrelation. If the cross-correlation has harmonic content at some frequency $f$, one can interpret as the signals being correlated at that frequency. Thus, this is useful in the quantitative comparison of power spectra of different signals. It is important to note spectral coherence measure assumes stationary processes.

Spectral coherence is calculated between $N$ trials of two time series  $x(t)$ and $y(t)$. We use the following modified formula for spectral coherence, as mentioned in~\cite{Lowet2016} which minimizes the effect of amplitude correlation.
\begin{equation}
C_{xy}(f) = \left| \frac{1}{N}  \sum_{n=1}^{N}\frac{G_{xy}(f)} {\sqrt{G_{xx}(f) G_{yy}(f)}} \right| 
\label{coh-eqn}
\end{equation}

\subsection{Mutual Information}
\label{sec:26}
Mutual information is an information-theoretic measure that can quantify non-linear dependencies between systems, unlike linear cross-correlation. It quantifies the amount of information about one system obtained by knowing about the other system \cite{kreuz2013synchronization} \cite{tzannes1973mutual}. Kullback-Leibler divergence (KLD) is a measure of the non-symmetric difference between two probability distributions $P$ and $Q$. Mutual information is defined as KLD between joint probability distribution of two random variables $p_{xy}(i,j)$ and product of their marginal distribution $p_{x}(i) \cdot p_{y}(j)$  \cite{kreuz2013synchronization} \cite{tzannes1973mutual}, as follows,

 \begin{equation}
 I(X,Y) = - \sum_{i=1}^{M_x} \sum_{j=1}^{M_y} p_{xy}(i,j) log \left(\frac{p_{xy}(i,j)}{p_x(i) \cdot p_y(j)}\right)
  \label{mi-eq}
\end{equation}

where $p_{x}(i), i = 1,...,M_x$ represents the probabilities of the $i$-th state in $X$ space and $M_x$ denotes the number of states etc.

Classical approaches for estimating mutual information includes finite-sized binning or bins with adjustable sizes. Alternatively, Kernel Density Estimate (KDE) approach is a non-parametric method for probability densities, and is said to be better than histogram or binning approaches; for example it has a better mean-squared error rate of convergence of the estimate to the underlying distribution (see~\cite{moon1995estimation} for details). In SyncBox, we have used the KDE approch to calculate Mutual Information.

\subsection{Phase Slip and Polar Histogram}
\label{sec:22}
In addition to the above measures, Phase Slip between the responses of neural population is a good visual indicator of their phase relations~\cite{Notbohm2016,Rosenblum2001}. Using the phase relations data extracted using the Hilbert Transform for the afore-mentioned measures, the unwrapped phase values are plotted against time. A plateau in the phase slip plots indicates phase synchrony, while a ramp indicates out-of-phase where the degree of mismatch is indicated by the slope of the ramp.

Synchrony in the phase relations of two signals are also well visualised by Polar Histograms~\cite{Notbohm2016}. A tight cluster indicates high phase locking, while a noisy relation is indicated by uniformly distributed bins.

\subsection{The biologically inspired population neural network}
\label{sec:27}
\begin{figure}
    \centering
    \includegraphics[width=2.5in]{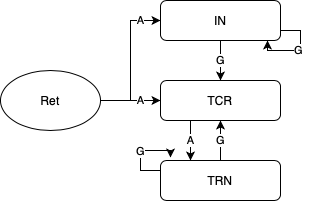}
    \caption{The biologically-inspired neural mass model of the Lateral Geniculate Nucleus (LGN --- visual thalamus). The three neural populations of the LGN are the thalamocortical relay (TCR), thalamic reticular nucleus (TRN), and local interneurons (IN). External input to the LGN is from the retinal neurons (Ret). The arrows represent synaptic connections and the label indicate the neurotransmitter at the synapse; `A' indicates AMPA-based and `G' indicates GABA(A)-based neurotransmission}
    \label{fig:lgn}
\end{figure}

The existing model of the visual thalamus, referred to as the Lateral Geniculate Nucleus (LGN), has three neuronal populations viz. Relay cells (TCR), Reticular Nucleus (TRN) and Interneurons (IN). In an earlier study, we have observed that removing the IN feed-forward connectivity in the network allows for synchrony to set in between the relay cell (excitatory) and reticular neucleus (inhibitory)populations. The synchrony between these two LGN populations have been studied since long in context to understanding their coupled dynamics in neurological disorders. (For details on the LGN model and its dynamics, reader may please refer to~\cite{bsb_springerbookchap_2016}). We have tested SyncBox under the presence and absence of IN. The phase locking behaviour indicated in the network conforms to that observed in our previous studies on the model.

\subsection{Simulation Methods}
\label{sec:28}
All algorithms presented in SyncBox are simulated in MATLAB\cite{MATLAB:2020b}. Test signals are model responses simulated at 1 msec resolution, filtered with Butterworth bandpass filter between 1 and 200 Hz. Sampling frequency for frequency domain computations was 1000 Hz. The model was simulated for a total of 40 seconds, and the outputs were then extracted between 1 and 30 seconds. The impulse input frequencies are varied from 1 to 30 Hz; all inputs are mixed with uniform noise at a mean -65 mV and a standard deviation of 2 mV.

\section{Results and Discussion}
\label{sec:3}
\begin{figure}
    \centering
    \begin{subfigure}{0.5\textwidth}
      \centering
      \includegraphics[width=\textwidth]{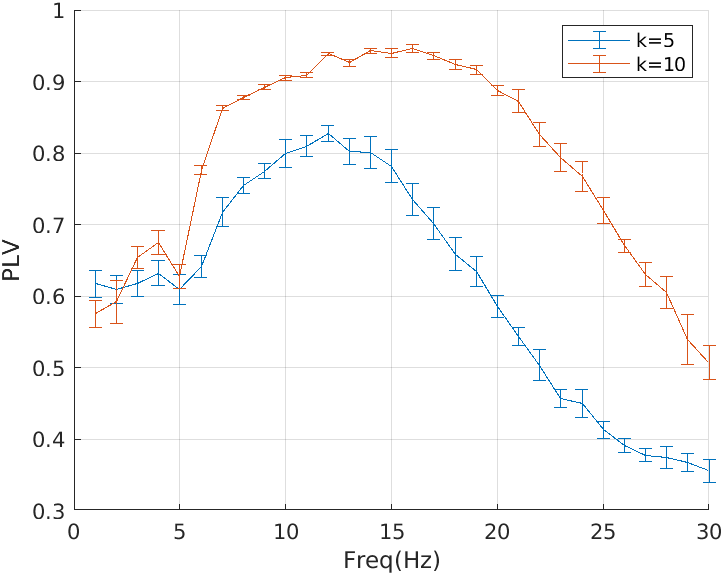}
      \caption{PLV}
      \label{fig:PLV_withIN}
    \end{subfigure}%
    \begin{subfigure}{0.5\textwidth}
      \centering
      \includegraphics[width=\textwidth]{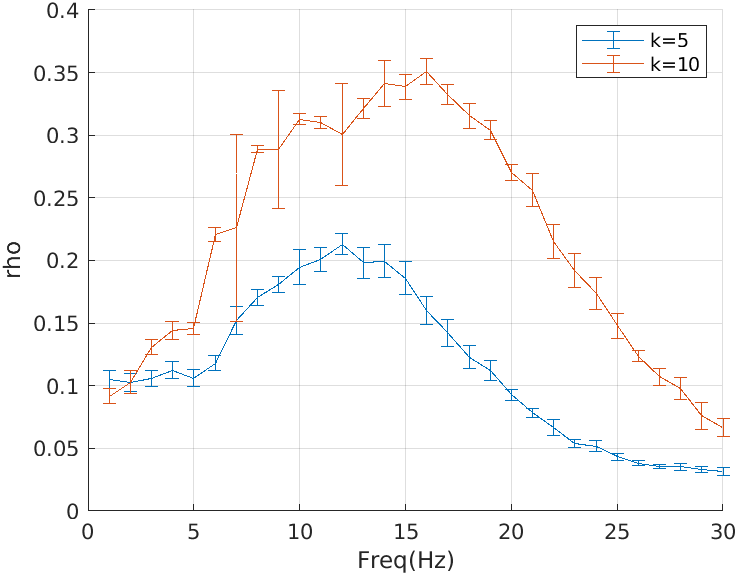}
      \caption{NSE}
      \label{fig:Rho_withIN}
    \end{subfigure}\\
    \begin{subfigure}{0.5\textwidth}
      \centering
      \includegraphics[width=\textwidth]{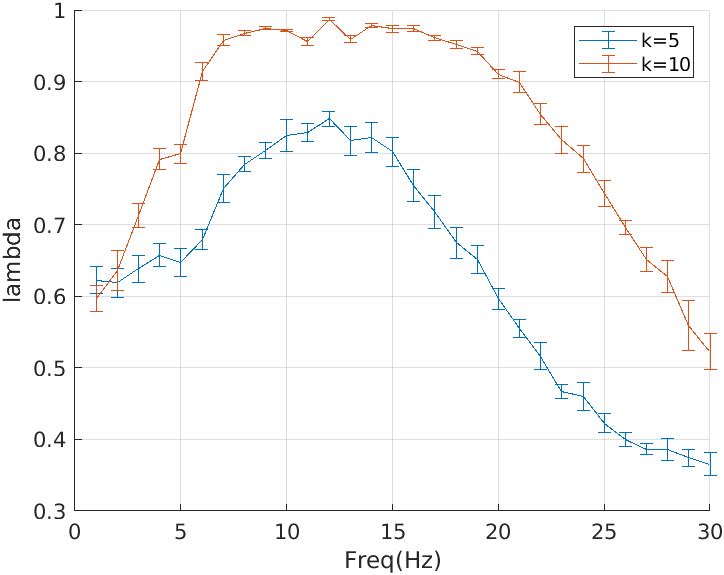}
      \caption{LSI}
      \label{fig:Lambda_withIN}
    \end{subfigure}%
    \begin{subfigure}{0.5\textwidth}
      \centering
      \includegraphics[width=\textwidth]{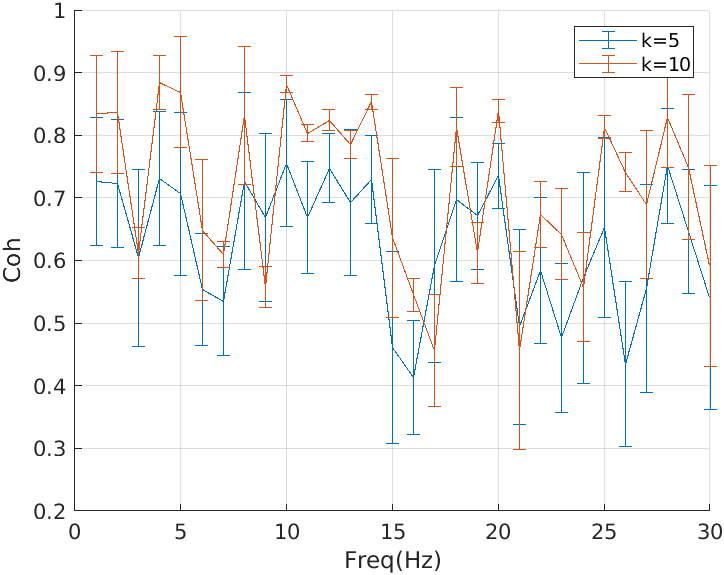}
      \caption{Spectral Coherence}
      \label{fig:Coh_withIN}
    \end{subfigure}%
    \caption{Quantitative measure of phase synchronisation in the LGN neural network, specifically between TCR and TRN populations, using (a) Phase Locking Value, (b) Normalised Shannon Entropy, (c) Lambda Index measure, and (d) Spectral Coherence.}
    \label{fig:lineplots}
\end{figure}
At first, we have tested our implementation of the SyncBox on simple coupled oscillators as indicated in~\cite{Lowet2016}. Next, we apply SyncBox on the LGN network response, and our observations are presented in this section. 

In Fig.~\ref{fig:lineplots}, we present a comparative study of synchronization in the model with a high and a low signal-to-noise ratio (amplitude of the input impulse as 5 and 10 respectively). As observed in~\cite{bsb_springerbookchap_2016},
we note that PLVs between TCR and TRN time-series responses are high between frequencies 5~--~20. The NSE and LSI also confirm this observation. However, the Spectral coherence plots are jittery. Interestingly, Lowet et al~\cite{Lowet2016} showed that within a large parameter range, the spectral coherence measure deviated substantially from the expected phase locking. They showed that spectral coherence did not converge to the expected value with increasing Signal-to-Noise-Ratio (SNR) and failed particularly when synchronization was partial or intermittent, which is expected to be the most likely state for neural synchronization. One of the reasons is that spectral coherence measure assumes the stationarity in the signals whereas PLV do not rely on the stationarity of the signals and are therefore much more accurate, reliable and broadly applicable. Overall, our results showing Spectral Coherence to be a less reliable measure of phase synchrony agrees with those of Lowet et al. \cite{Lowet2016}.

\begin{figure}
    \centering
    \includegraphics[width=0.7\textwidth]{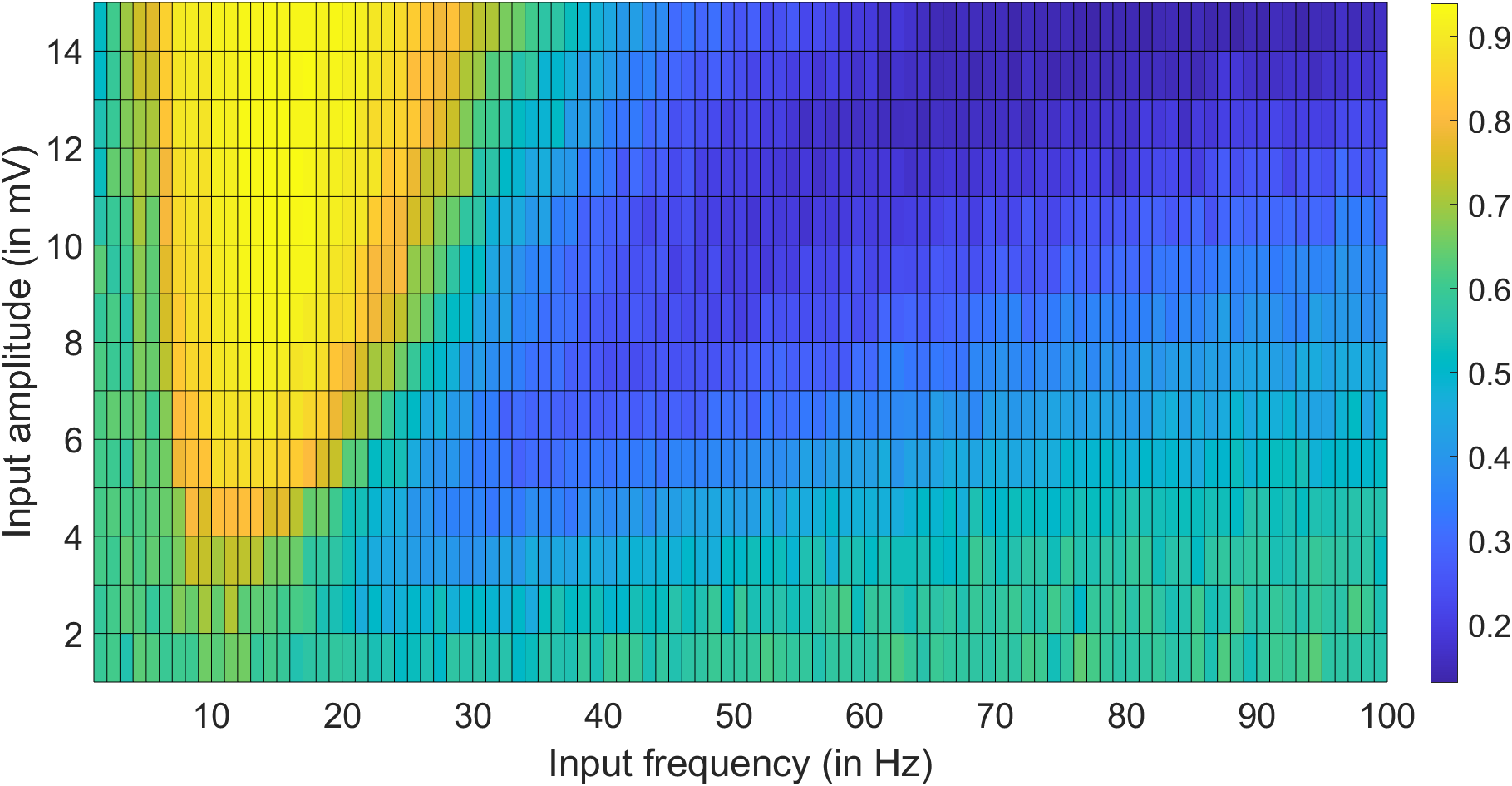}
    \caption{(a) Arnold Tongue formation of phase relations between TCR and TRN populations of the LGN model when stimulated with rhythmic impulse inputs with varying frequencies and amplitudes. Each pixel in the plot indicates the mean of the PLV between the two population response time-series at a certain input frequency and amplitude.}
    \label{fig:arnoldT}
\end{figure}
\begin{figure}
    \centering
    \includegraphics[width=0.7\textwidth]{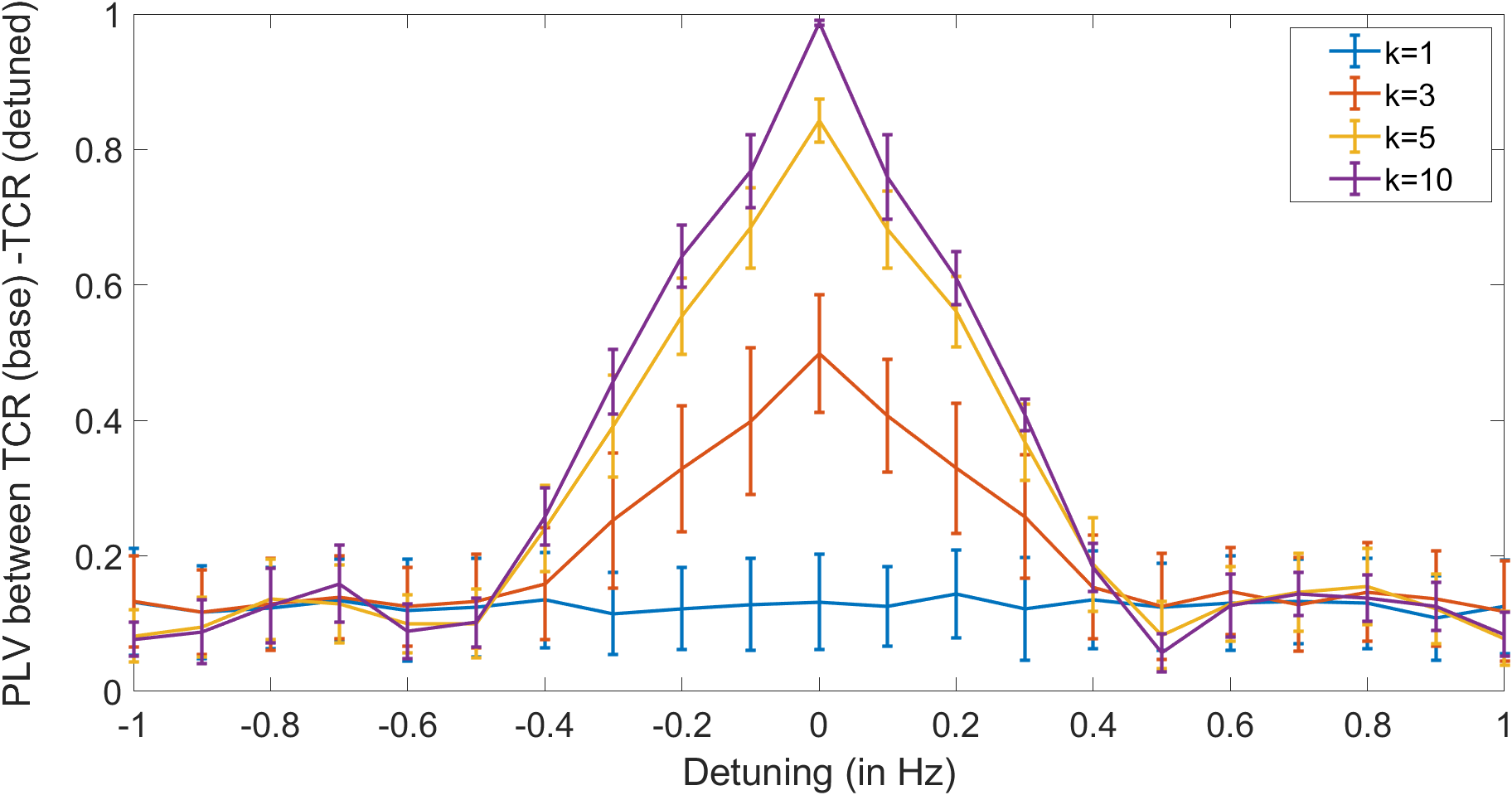}
    \caption{Phase locking between base frequency output and detuned frequency output with respect to detuned input at varied amplitudes.}
    \label{fig:detuningTCR}
\end{figure}
As indicated in Sec.~\ref{sec:21}, the mean of the PLVs between two neural populations, corresponding to varying input frequency and amplitude, can be plotted as a 2-D grid. The degree of synchrony often increases with increasing input amplitude and can be observed as an Arnold Tongue formation. Figure~\ref{fig:arnoldT} shows the Arnold Tongue formation between the TCR and TRN population responses with input impulse train frequencies varied over 1 to 100 Hz, and with impulse amplitude varied over 1 to 15 mV. The line plots in Fig.~\ref{fig:lineplots}(a) also conform to the Arnold Tongue formation in the circuit. The 2-D plot gives a global map of regions of phase synchronization in the population network for different input amplitudes and input frequencies.

Another visualisation technique to understand phase synchrony between neural populations is to drive them at `detuned' frequency with reference to a base frequency, and observing the PLV. The degree of synchronisation between two coupled system under study will depend on the difference between their intrinsic frequencies referred to as detuning ($\Delta \omega$), and their coupling strength $\kappa$. The phase-locking behaviour can be observed by simultaneously varying $\Delta \omega$ and $\kappa$. In Fig.~\ref{fig:detuningTCR}, we have shown the PLV over shorter durations (2 second intervals) between the output timeseries of TCR populations the LGN network at base input frequency ($10$ Hz) and a detuned input frequency ($10 \pm \Delta \omega$ Hz). We vary the amplitude of input signal which thereby increases SNR, thus has a similar effect to that of varying the coupling strength between two oscillators . The input impulse frequency to one TCR population is maintained at 10 Hz and is noise free, while the input impulse frequency of the second TCR population is varied as $ 9 \leq (10 \pm \Delta \omega) \leq 11$ (Hz), $\Delta \omega \geq 0.1$ (Hz) and is noisy. This is done for varying input amplitudes $\kappa$. As expected, with higher $\kappa$, the PLV is higher. However, the width of the `inverted tongue' is the same irrespective of $\kappa$. Readers may note that we have only used $1:1$ PLV measure in this study; a full exploration of the $m:n$ space may show further interesting dynamics and is left as future work. For longer durations of 30 seconds, we observe no sustained synchronization in this set up. Future work can include a study of synchronization when the two LGN models are coupled with biologically inspired connections in the differential equation based model, where one can vary these coupling strengths directly instead of varying the SNR and we can expect to see sustained phase-locking in such scenarios, as seen in simpler Izhikevich based coupled networks \cite{Lowet2016}.

\begin{figure}
    \centering
    \begin{subfigure}{0.5\textwidth}
      \centering
      \includegraphics[width=0.9\textwidth]{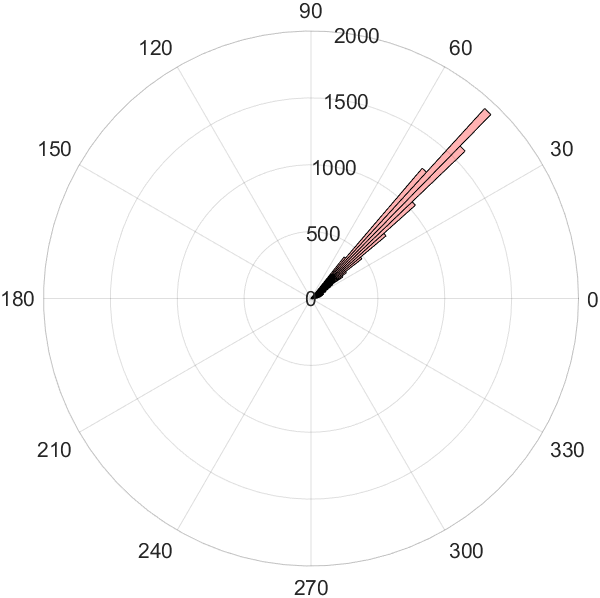}
      \caption{Polar histogram for high synchrony}
      \label{fig:Polarhist_high_sync}
    \end{subfigure}%
    \begin{subfigure}{0.5\textwidth}
      \centering
      \includegraphics[width=0.9\textwidth]{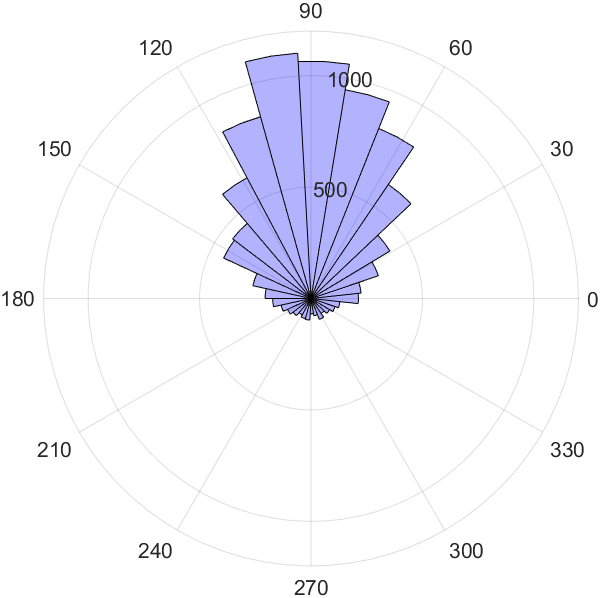}
      \caption{Polar histogram for low synchrony}
      \label{fig:Polarhist_low_sync}
    \end{subfigure}\\ <<
    \begin{subfigure}{\textwidth}
      \flushleft
      \includegraphics[scale=0.35]{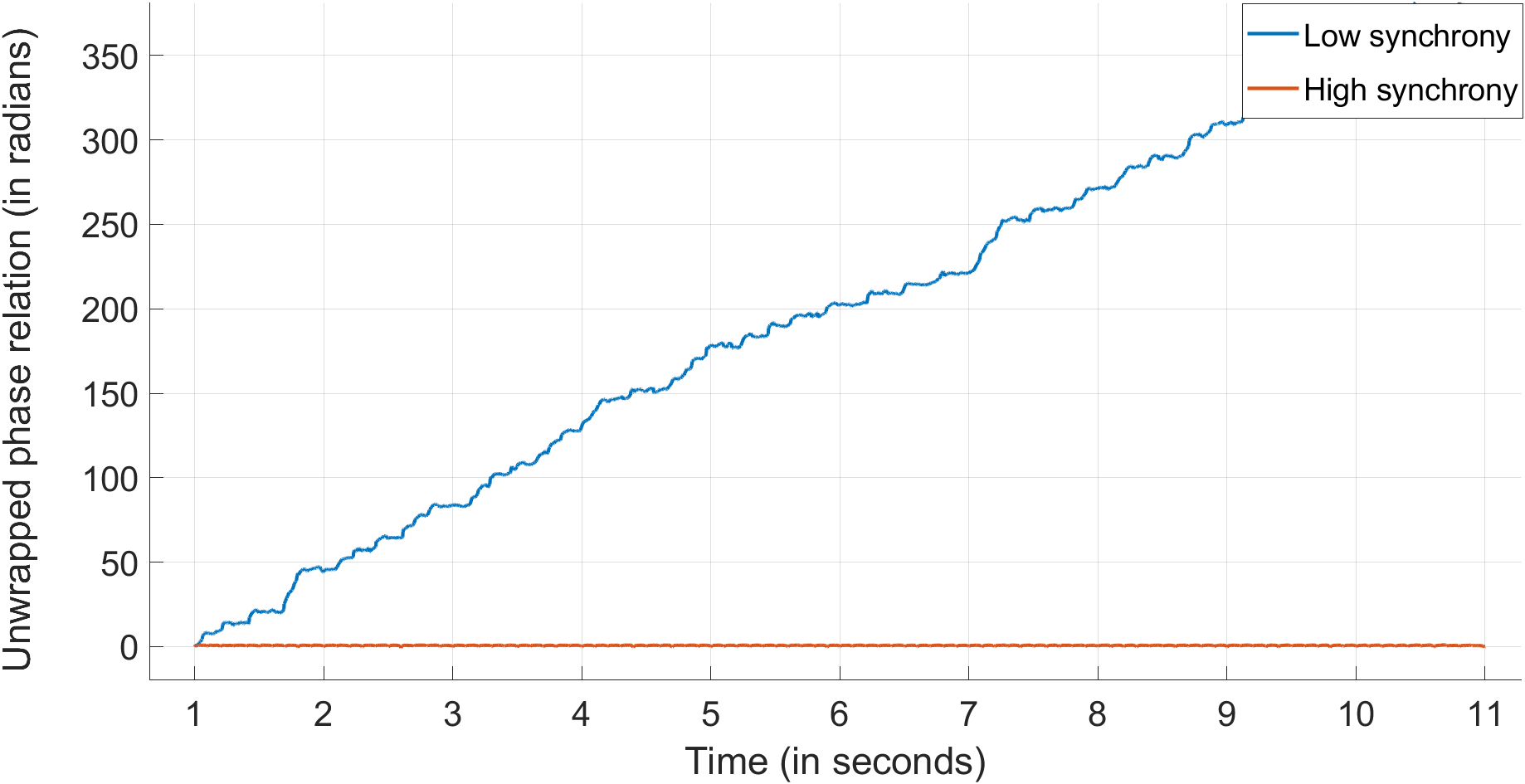}
      \caption{Comparison of phase slips for low and high synchrony}
      \label{fig:Phase_slips}
    \end{subfigure}
    \caption{Polar Histogram showing (a) high and (b) low phase synchrony between the TCR and TRN populations in the LGN model. The condition of high synchrony in (a) is with IN disconnected from LGN circuit, input impulse frequency and amplitude of 10 Hz and 10 mV respectively, and high SNR. The condition of low synchrony in (b) is with full circuit simulation, input impulse frequency and amplitude of 2 Hz and 5 mV respectively, and low SNR. (c) The Phase slip plots corresponding to the states of high synchrony (red, flat) and low synchrony (blue, ramp).}
    \label{fig:visuals}
\end{figure}
Lastly, we demonstrate phase relation between TCR and TRN populations using Phase slip plots (Fig.~\ref{fig:visuals}) and Polar Histogram (Figures~\ref{fig:Polarhist_low_sync},~\ref{fig:Polarhist_high_sync}). As indicated in Sec.~\ref{sec:27}, the LGN model is simulated as full circuit, and then with the IN removed, and the phase relation between the TCR and TRN is captured. Narrow clustered formation in Fig.~\ref{fig:Polarhist_high_sync} indicate high phase synchrony when model is simulated with IN disconnected and high SNR~---impulse input at 10 Hz and at an amplitude of 10 mV. In contrast, with IN connected in the circuit, and with a low SNR, impulse input at 2 Hz and at an amplitude of 5 mV, the synchrony between TCR and TRN is lower and the Polar Histogram has a wider spread of the bins as shown in Fig.~\ref{fig:Polarhist_high_sync}. The phase slip plots in Fig.~\ref{fig:Phase_slips} confirm this observation---~plot is flat for the high SNR, high synchrony case, and is a ramp for low SNR, low synchrony.

\section{Conclusion and Future work}
We have been working on understanding phase-locking, and phase synchronisation in an existing biologically inspired population neural network, commonly referred to as a neural mass model, of the visual thalamus. Towards this, we have collated a set of tools implemented on Matlab, SyncBox. The algorithmic implementation in SyncBox are informed by those in~\cite{kreuz2013synchronization} (Mutual Information), \cite{Kraskov2004} (Phase Locking Value, Spectral Coherence), \cite{Notbohm2016} \cite{Rosenblum2001} (Normalised Shannon Entropy, Lambda Synchronisation Index). We have also added visualising phase synchrony using Arnold Tongue plot, Phase slip plot and Polar Histogram as indicated in~\cite{Notbohm2016}. We have tested the SyncBox on our model, under two known parameterised condition as presented in~\cite{bsb_springerbookchap_2016}. The observed synchronisation measures agree with those observed qualitatively in time-series plots.

Encouraged by our positive test results above, we are now in the process of using the SyncBox measures in an enhanced neural mass model of the visual pathway. However, we understand that we could not include other useful measures of phase synchronisation (e.g. phase transfer entropy) in the toolbox due to time constraints. Furthermore, and although a part of our toolbox, we have not shown the Mutual Information plots as our study on this measure is still in progress; we intend to demonstrate this in upcoming publications.  As future work, we endeavour to improve upon the currently available repertoire of tools in SyncBox, which will also be made freely availables for use by the larger scientific community. 
\label{sec:4}

\bibliographystyle{splncs04}
\bibliography{bibliography}

\end{document}